\documentclass[prb,showpacs,amsmath,amssymb]{revtex4}

\usepackage{graphicx}
\usepackage{dcolumn}
\usepackage{bm}

\begin{document}
\title{Superlattice properties of carbon nanotubes in a transverse 
electric field}
\author{O.~V.~Kibis}
\email{Oleg.Kibis@nstu.ru}
\affiliation{Department of Applied and Theoretical Physics, 
Novosibirsk State Technical University, Novosibirsk 630092, Russia}
\author{D.~G.~W.~Parfitt and M.~E.~Portnoi}
\affiliation{School of Physics, University of Exeter, Stocker
Road, Exeter EX4 4QL, United Kingdom}
\begin{abstract}
Electron motion in a $(n,1)$ carbon nanotube is shown to
correspond to a de Broglie wave propagating along a helical line
on the nanotube wall. This helical motion leads to periodicity of
the electron potential energy in the presence of an electric
field normal to the nanotube axis. The period of this potential
is proportional to the nanotube radius and is greater than the
interatomic distance in the nanotube. As a result, the behavior
of an electron in a $(n,1)$ nanotube subject to a transverse
electric field is similar to that in a semiconductor
superlattice. In particular, Bragg scattering of electrons from
the long-range periodic potential results in the opening of gaps
in the energy spectrum of the nanotube. Modification of the
bandstructure is shown to be significant for experimentally
attainable electric fields, which raises the possibility of 
applying this effect to novel nanoelectronic devices.
\end{abstract}
\pacs{73.22.-f, 73.63.Fg, 78.67.Pt}
\maketitle

\section{Introduction}

Carbon nanotubes (CNTs) are cylindrical molecules with nanometer
diameter and micrometer length. Since the discovery of CNTs just over
a decade ago,\cite{Iijima} their unique electronic and structural 
properties have aroused great excitement in the scientific community 
and promise a broad range of applications. 
Significant theoretical effort has been applied to develop refined 
models of the electronic structure of carbon nanotubes, as well as 
their optical and transport properties, although even a simple 
tight-binding model\cite{Dresselhaus} yielding analytic solutions is 
sufficient to elucidate key nanotube features (e.g. whether a CNT of 
given structure will exhibit metallic or semiconducting properties). 
In this paper we apply such a model to a particular type of 
single-wall CNT, a so-called $(n,1)$ nanotube. 
In Sec.~\ref{noField} we show that for such a 
CNT the electron motion corresponds to a de Broglie wave propagating 
along a helical line. The theoretical treatment of this type of CNT in 
an electric field perpendicular to the nanotube axis (transverse 
electric field) can be reduced to a one-dimensional superlattice 
problem (see Sec.~\ref{field}). Such superlattice behavior of 
current-carrying electrons suggests the application of CNTs to 
the development of novel carbon nanotube-based devices.

\section{Energy spectrum of $(n,1)$ nanotubes}\label{noField}

\begin{figure}
\includegraphics[width=0.9\textwidth]{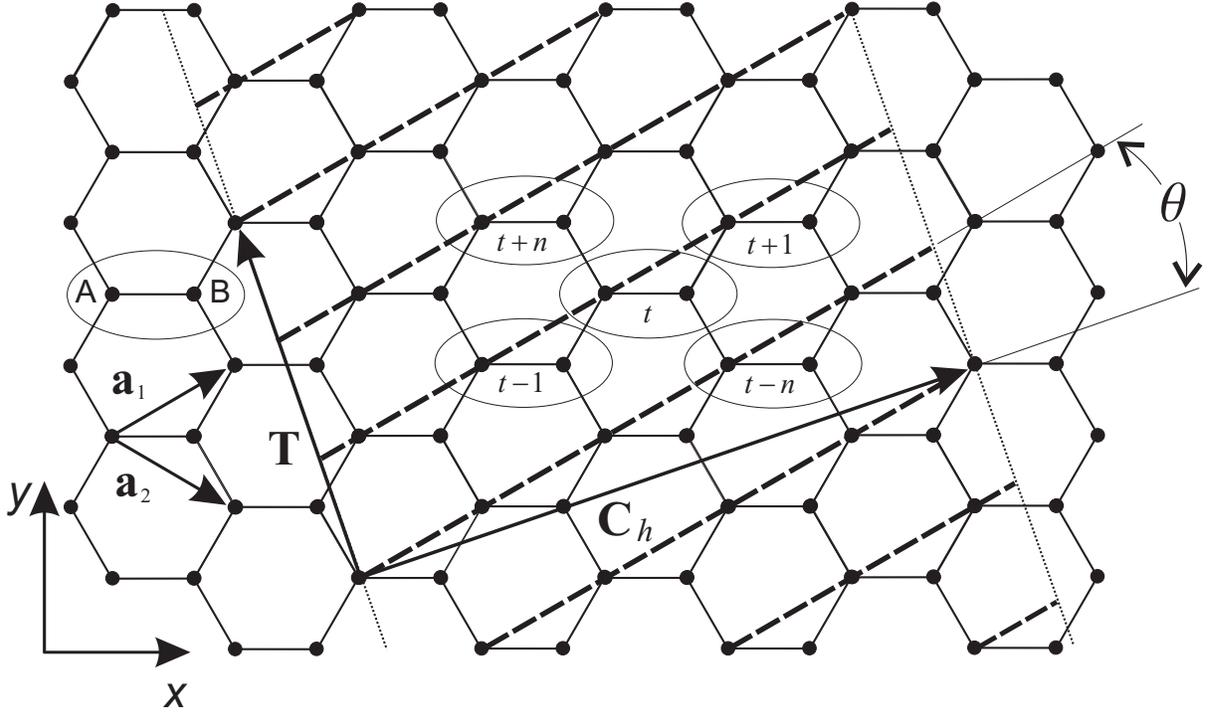}
\caption{The unrolled graphite sheet. By connecting the head and
tail of the chiral vector $\mathbf{C}_h$ we can construct, for
example, a $(4,1)$ carbon nanotube. The dashed lines will then
form a helical line on the nanotube wall.} 
\label{Fig1}
\end{figure}

A single-wall carbon nanotube may be considered as a single
graphite sheet rolled into a cylinder. The electronic energy
spectrum of the CNT is therefore intimately related to the energy
spectrum $\varepsilon_{g2D}(\mathbf{k})$ of a two-dimensional (2D)
graphite sheet, which can be written in the tight-binding
approximation as\cite{Dresselhaus}:
\begin{equation}\label{eg2d}
\varepsilon_{g2D}(\mathbf{k})=\pm\gamma_0\left|\exp\left(
\frac{ik_xa}{\sqrt{3}}\right)+2\exp\left(-\frac{ik_xa}{2\sqrt{3}}
\right)\cos\left(\frac{k_ya}{2}\right)\right|,
\end{equation}
where $k_x$ and $k_y$ are the electron wave vector components in
the graphite sheet plane along the $x$ and $y$ axes, respectively
(see Fig.~\ref{Fig1}). 
In the energy spectrum \eqref{eg2d}, the
plus and minus signs correspond to the conduction and valence
bands, respectively, $\gamma_0\approx 3$ eV is the transfer
integral between $\pi$-orbitals of neighboring carbon atoms, and
the lattice constant
$a=|\mathbf{a}_1|=|\mathbf{a}_2|=\sqrt{3}\times
a_{\mbox{c-c}}=2.46$ {\AA}, where $\mathbf{a}_1$ and
$\mathbf{a}_2$ are the 2D basis vectors and $a_{\mbox{c-c}}=1.42$
{\AA} is the interatomic distance in graphite. The way in which
the 2D graphite sheet is rolled up to form the CNT can be
described by two vectors, the translation vector $\mathbf{T}$ and
the chiral vector $\mathbf{C}_h$ (see Fig.~\ref{Fig1}). The
chiral vector $\mathbf{C}_h$ can be expressed in terms of the 2D
basis vectors of the unrolled graphite sheet as
$\mathbf{C}_h=n\mathbf{a}_1+ m\mathbf{a}_2$, where the pair of 
integers $(n,m)$ is used as a standard notation\cite{Dresselhaus} 
for a CNT of given crystal structure. To obtain the electronic energy 
spectrum of the $(n,m)$ CNT, we begin by expressing the wave vector 
$\mathbf{k}$ in terms of components along $\mathbf{T}$ and 
$\mathbf{C}_h$ as $\mathbf{k}=k_\|\mathbf{T}/T+
k_\perp\mathbf{C}_h/C_h$, where $k_\|$ and $k_\perp$ are 
subject to the following constraints: $-\pi/T<k_\|\leq\pi/T$ and
$k_\perp=2\pi l/C_h$ $(l=0,1,2,\ldots ,N-1)$.
The integer $l$ represents the electron angular momentum along the 
nanotube axis and
\begin{equation}\label{Ndef}
N=\frac{2(n^2+m^2+nm)}{d_R},
\end{equation}
is the number of elementary atomic cells consisting of two carbon
atoms ($A$, $B$) per area
$\left|\mathbf{C}_h\times\mathbf{T}\right|$. The number $d_R$
appearing in Eq.~\eqref{Ndef} is the greatest common divisor of the
two integers ($2n+m$, $2m+n$).
The lengths of the chiral vector and translation vector are given by
$C_h=a\sqrt{n^2+m^2+nm}$ and $T=\sqrt{3}C_h/d_R$, respectively.

\begin{figure}
\includegraphics[width=0.9\textwidth]{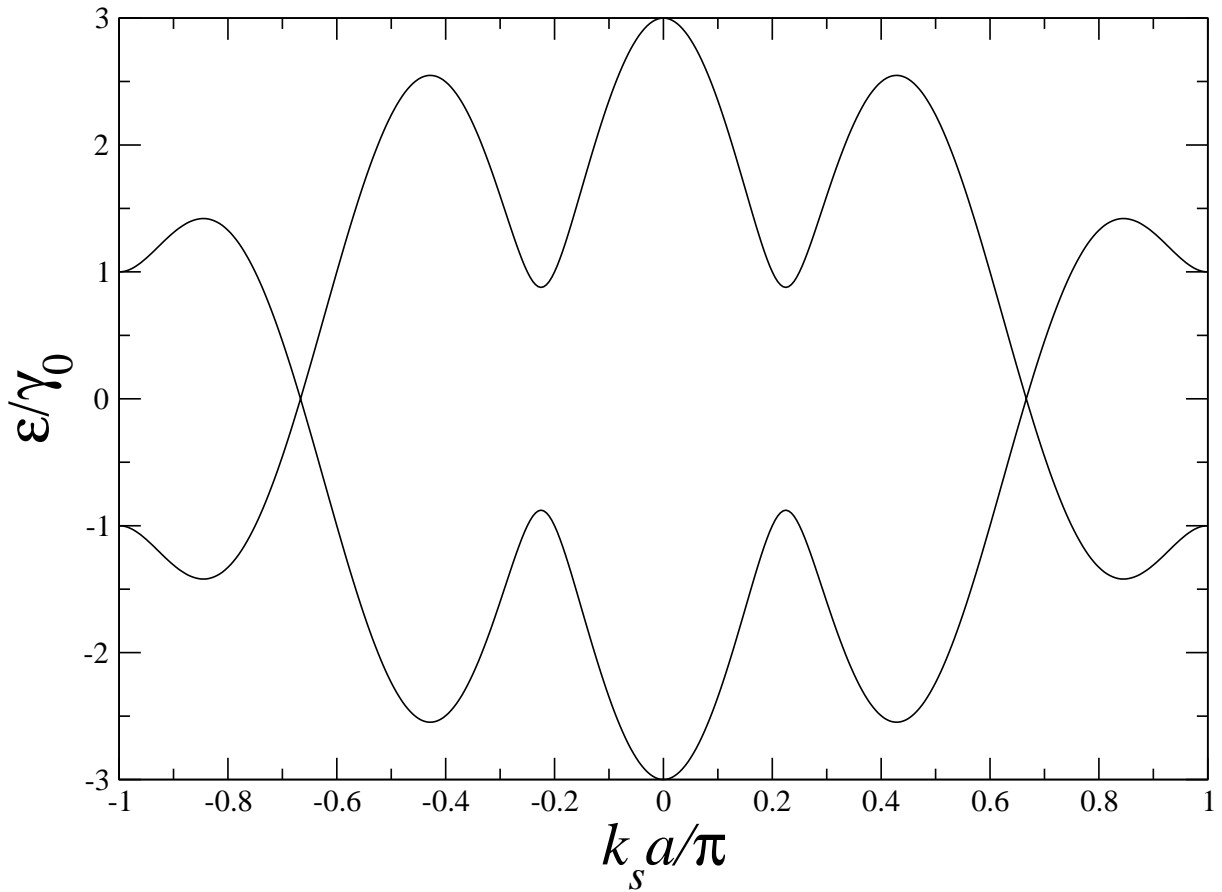}
\caption{Electron energy spectrum of a metallic $(4,1)$ CNT
as a function of the wave number $k_s$ along a helical line on 
a nanotube wall.}
\label{Fig4_1noField}
\end{figure}

The energy spectrum of a $(n,m)$ CNT can be obtained by expressing
$k_x$ and $k_y$ in terms of $k_\|$ and $k_\perp$, and substituting
them in Eq.~\eqref{eg2d}, thus yielding
\begin{equation}\label{neweg2d}
\varepsilon=\pm\gamma_0\left|\exp\left[
\frac{i\sqrt{3}a}{2}\left(k_\|\cos\theta-k_\perp\sin\theta\right)
\right] +2\cos\left(\frac{k_sa}{2}\right)\right|,
\end{equation}
where we have introduced the new parameter 
$k_s=k_\perp\cos\theta+k_\|\sin\theta$, and the chiral angle 
$\theta$ $(|\theta|\le\pi/6)$ shown in Fig.~\ref{Fig1}. Taking into 
account that
\begin{equation}\label{costheta}
\cos\theta =\frac{2n+m}{2\sqrt{n^2+m^2+nm}},\quad
\sin\theta =\frac{\sqrt{3}m}{2\sqrt{n^2+m^2+nm}},
\end{equation}
we have, for $m\neq 0$, the equation
\begin{equation}\label{mnoteq}
\sqrt{3}(k_\|\cos\theta-k_\perp\sin\theta)a=[(2n+m)k_sa-2k_\perp
C_h]/m.
\end{equation}
Substituting Eqs.~\eqref{costheta} and \eqref{mnoteq} into
Eq.~\eqref{neweg2d} we obtain
\begin{equation}
\varepsilon=\pm\gamma_0\left|\exp\left[i\left(
\frac{2n+m}{2m}k_sa-\frac{k_\perp
C_h}{m}\right)\right]+2\cos\left(\frac{k_sa}{2}\right)\right|,
\end{equation}
which, together with the constraint $k_\perp=2\pi l/C_h$, yields 
an electron energy spectrum of the form
\begin{equation}\label{epsi}
\varepsilon=\pm\gamma_0\left[1+4\cos\left(\frac{k_sa}{2}\right)
\cos\left(\frac{2n+m}{2m}k_sa
-\frac{2\pi l}{m}\right)+4\cos^2\left(\frac{k_sa}{2}\right)
\right]^{1/2}.
\end{equation}
For $m=1$, Eq.~\eqref{epsi} becomes independent of $l$, and we
obtain the electron energy spectrum of a $(n,1)$ CNT in the form
\begin{equation}\label{ejks}
\varepsilon_j(k_s)=(-1)^j\gamma_0\left[1+8\cos\left(
\frac{n+1}{2}k_sa\right)\cos\left(\frac{nk_sa}{2}\right)
\cos\left(\frac{k_sa}{2}\right)\right]^{1/2},
\end{equation}
where $j=1,2$ correspond to the valence and conduction bands,
respectively. It should be noted that the spectrum \eqref{ejks}
depends on the parameter $k_s$ alone, in contrast to the general
case of a $(n,m)$ CNT, for which the electron energy spectrum
depends on two parameters ($k_\|$ and $k_\perp$ are
conventionally used). This peculiarity of a $(n,1)$ CNT is a
consequence of its special crystal symmetry: the $(n,1)$ CNT
lattice can be obtained by translation of an elementary two-atom
cell along a helical line on the nanotube wall (see
Fig.~\ref{Fig1}). As a result, the parameter $k_s$ has the
meaning of an electron wave vector along the helical line, and so
any possible electron motion in a $(n,1)$ CNT can be described by
a de Broglie wave propagating along such a line. 
Thus, $(n,1)$ CNTs represent a previously overlooked distinctive
class of nanotubes, which may be termed `helical' nanotubes.
The electron energy spectrum of a (4,1) CNT as a function of the 
helical wave number $k_s$ is shown in Fig.~\ref{Fig4_1noField}.
The band gap for this natotube closes at $k_sa=2\pi /3$, and it can 
be shown that the same is true for all metallic $(n,1)$ nanotubes.

\section{Helical nanotubes in a transverse electric field}\label{field}

Both descriptions of the energy spectrum of a $(n,1)$ CNT --- by 
two parameters, $k_\|$ and $k_\perp$, or a single parameter $k_s$
--- are physically equivalent. However, the second description is
more convenient for studies of electron processes determined by
the above-mentioned helical symmetry of electron motion, and
allows one to discover new physical effects (e.g. the
electron-electron interaction should be strongly modified for
helical one-dimensional motion\cite{KibisPLA}). We shall now show
that such helical symmetry results in superlattice behavior of a
$(n,1)$ CNT in the presence of an electric field oriented
perpendicular to the nanotube axis (a transverse electric field).

The potential energy of an electron on a helix subject to a
transverse electric field takes the form
\begin{equation}\label{pe}
U=eER\cos\left(\frac{2\pi s}{l_0}\right),
\end{equation}
where $e$ is the electron charge, $E$ is the electric field
strength, $R=C_h/2\pi$ is the radius of the CNT, $s$ is the
electron coordinate along the above-mentioned helical line,
\begin{equation}\label{lzero}
l_0=\frac{2\pi R}{\cos\theta}=\frac{2a(n^2+n+1)}{2n+1},
\end{equation}
is the length of a single coil of the helix, and the electric
potential is assumed to be zero at the axis of the CNT. The
potential energy \eqref{pe} is periodic in the electron coordinate
$s$ along the helical line and the period of the potential is equal
to $l_0$. Since this period \eqref{lzero} is proportional to the
CNT radius $R$ and is greater than the interatomic distance
$a_{\mbox{c-c}}$, the CNT assumes typical superlattice properties.
In particular, Bragg reflection of electron waves with wave
vectors $k_s=\pm \pi/l_0$ results in energy splitting within the
conduction and valence bands of the CNT. We shall now study this
effect in more detail.

In the framework of the tight-binding model,\cite{Dresselhaus}
considering only three nearest neighbors to each atom, the wave
functions for electron states with corresponding energies 
\eqref{ejks} can be written as
\begin{equation}\label{tight}
\psi_j(k_s)=\frac{1}{\sqrt{2M}}\sum_t\left[\psi^{(A)}_t+(-1)^j
\frac{h^\ast(k_s)}{|h(k_s)|}\psi^{(B)}_t\right]\exp(ik_sta),
\end{equation}
where $M$ is the total number of two-atom cells in the CNT,
$\psi^{(A)}_t$ and $\psi^{(B)}_t$ are $\pi$-orbital wave functions
for the two carbon atoms $A$ and $B$, respectively, $t$ is the
number along the helical line for an elementary cell consisting of
these two atoms (see Fig.~\ref{Fig1}), and 
$h(k_s)=1+\exp(-ik_sa)+\exp(ink_sa)$.
The value of the potential energy $U$ in the external electric
field at the position of a particular atom of the CNT depends on
the angle between the electric field vector and the vector normal
to the nanotube axis which passes through this atom.
As a consequence, the coordinate of atom $A$ in cell number $t$
along the helical line is
\begin{equation}\label{sdef}
s=at+\frac{l_0}{2\pi}\phi.
\end{equation}
The angle $\phi$ is defined in such a way that 
$R\cos\left[\phi+\pi(n+1)/(n^2+n+1)\right]$ is the coordinate in 
the direction of the electric field (with zero at the CNT axis) 
of atom $B$ in the cell with $t=0$. Using Eqs.~\eqref{tight} and 
\eqref{sdef}, we can write the matrix element of the potential 
energy \eqref{pe} as
\begin{equation}\label{matel}
\langle\psi_i(k_s^\prime)|U|\psi_j(k_s)\rangle =
V^+_{ij}\delta_{\cos(k_sa-k_s^\prime a+2\pi a/l_0),1}
+V^-_{ij}\delta_{\cos(k_sa-k_s^\prime a-2\pi a/l_0),1}\: ,
\end{equation}
where
\begin{equation}\label{Vij}
V^\pm_{ij}=\frac{eER}{4}\left[1+(2\delta_{ij}-1)
\frac{h(k_s^\prime)h^\ast(k_s)}{|h(k_s^\prime)h(k_s)|}
\exp\left(\pm i\frac{\pi(n+1)}{n^2+n+1}\right)\right] \exp(\pm
i\phi),
\end{equation}
and $\delta_{\alpha\beta}$ is the Kronecker delta.
In the derivation of Eqs.~\eqref{matel} and
\eqref{Vij} we have also assumed that the external electric field
$E$ is much less than the atomic field, i.e.
\begin{equation}\label{Eless}
E\ll\frac{\gamma_0}{ea}.
\end{equation}
This allows us to neglect any change in the atomic wave functions
$\psi^{(A)}_t$ and $\psi^{(B)}_t$ due to the field $E$, and we
take into account only the mixing of states \eqref{tight} by the
field. According to Eq.~\eqref{matel}, the field mixes only
electron states \eqref{tight} with wave vectors differing by
$2\pi/l_0$.
In this approximation, the exact wave function in the presence of
the electric field, $\psi_E(k_s)$, can be expressed as a
superposition of wave functions \eqref{tight} with $k_s$
shifted by integer numbers of $2\pi/l_0$:
\begin{equation}\label{phisum}
\psi_E(k_s)=\sum_{j=1}^2\sum_{\nu=0}^{\mu-1}b_{j\nu}
\psi_j\left(k_s+2\pi\nu /l_0\right).
\end{equation}
To ensure that in Eq.~\eqref{phisum} we sum only over
\emph{different} electron states, the parameter $\mu$ should be
the smallest integer defined by the condition
$\psi_j(k_s)=\psi_j\left(k_s+2\pi\mu/l_0\right)$. This condition,
together with the $2\pi/a$ periodicity of $\psi_j(k_s)$, implies
that $\mu/l_0=\beta/a$, where $\beta$ is the smallest integer for
which this equality is satisfied. Using Eq.~\eqref{lzero}
together with Eq.~\eqref{Ndef} one can obtain $\beta=(2n+1)/d_R$,
which yields $\mu=N$. This result has a transparent physical
interpretation, since the two closest carbon atoms equivalent
with respect to a translation parallel to the nanotube axis are
separated by a distance $Na$ along a helical line.

Substituting the wave function \eqref{phisum} into the
Schr\"{o}dinger equation with the potential energy \eqref{pe} we
obtain a system of equations for the coefficients $b_{j\nu}$
entering Eq.~\eqref{phisum}:
\begin{equation}\label{SE}
\left[\varepsilon_j\left(k_s+2\pi\nu /l_0\right)
-{\varepsilon_E(k_s)}\right]b_{j\nu}
+\sum_{i=1}^2\sum_{\nu^\prime=0}^{N-1}\langle\psi_j\left(k_s
+2\pi\nu /l_0\right) |U|\psi_i\left(k_s+2\pi\nu^\prime /
l_0\right)\rangle b_{i\nu^\prime}=0,
\end{equation}
where $\nu=0,1,2,\ldots ,N-1$, the index $j$ takes the value $1$
or $2$ for the valence and conduction bands, respectively, and
$\varepsilon_E(k_s)$ is the electron energy in the presence of
the transverse electric field. 

Let us consider the states
$k_s=-\pi/l_0$ and $\pi/l_0$ in the same CNT energy band, which
are at the boundaries of a Brillouin zone created by the
periodic `superlattice' potential \eqref{pe} of the external
field. One should expect the appearence of energy gaps at these
values of $k_s$ due to Bragg reflection of electron waves from the
superlattice potential. These states are separated by $2\pi/l_0$
and have the same energy, which means that they are strongly
mixed by the electric field. For these values of $k_s$ it can be
shown that the contributions to the sum in Eq.~\eqref{phisum}
from all other states can be neglected for sufficiently weak
fields, $E\ll\gamma_0a/(eR^2)$. As a result, the system of
equations \eqref{SE} is reduced to just two equations, from which
the energy of Bragg band splitting $\Delta\varepsilon$ is found
to be
\begin{equation}\label{Bragggeneral}
\Delta\varepsilon= 2\left|\langle\psi_j(-\pi/l_0)|
U|\psi_j(\pi/l_0)\rangle\right|\sim eER.
\end{equation}
Thus, even a small electric field results in a superlattice-like 
change of the electron energy spectrum in $(n,1)$ CNTs, with 
the appearance of Bragg energy gaps proportional to the field
amplitude $E$ and the nanotube radius $R$. Notably, this 
dependence of the Bragg gaps on the external field and radius 
applies to any helical quasi-one-dimensional nanostructure in a
transverse electric field: this generic feature arises from the 
symmetry of the nanostructure, and is independent of the 
parameters of the tight-binding model used to derive 
Eq.~\eqref{Bragggeneral}.
For example, it should be possible to observe a similar effect 
in recently fabricated InGaAs/GaAs and Si/SiGe semiconductor 
nanohelices.\cite{Prinz2000,Prinz2001}
 
It should be emphasized that for single-wall carbon nanotubes the 
discussed superlattice behavior is a unique feature of $(n,1)$ 
structures only. For the general case of a $(n,m)$ CNT with 
$m\neq 1$, the energy spectrum \eqref{epsi} depends on the quantum 
number $l$ in addition to $k_s$. As already mentioned, $l$ represents 
the projection of the electron
angular momentum on the nanotube axis, and it follows from the
corresponding selection rule that the transverse electric field
only mixes electron states with angular momentum $l$ and $l\pm1$.
For $m\neq1$, however, states with $l$ differing by one
correspond to different subbands, and in general have different
energies for $k_s=\pm\pi/l_0$, so that there is no Bragg
scattering between these states. The only effect of the electric
field, therefore, is to mix electron states with different
energies, which does not lead to noticeable modification of the
dispersion curves for weak electric fields.\cite{Rotkin}

\begin{figure}
\includegraphics[width=0.6\textwidth]{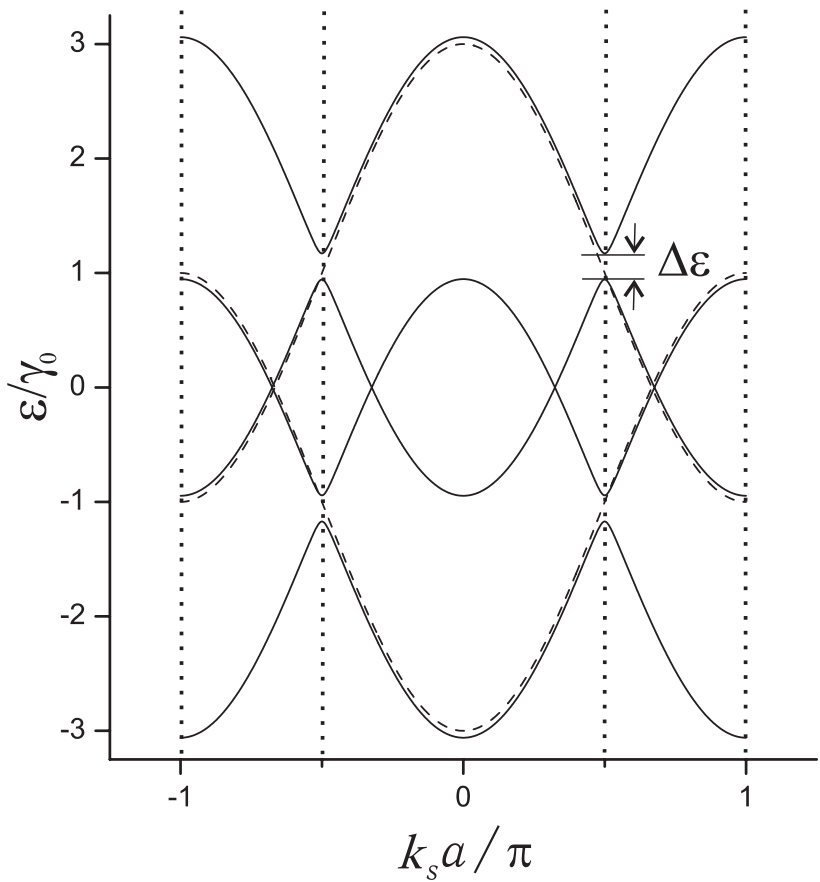}
\caption{Electron energy spectrum of a $(1,1)$ CNT in the presence
of a transverse electric field $E=\gamma_0/(ea_{\mbox{c-c}})$
with $\phi=0$ (solid lines) and without the electric field (dashed
lines). The inner pair of vertical dotted lines indicates the
first Brillouin zone boundary in the presence of the field,
whereas the outer pair corresponds to the first Brillouin zone
boundary without the field. $\Delta\varepsilon$ is the Bragg gap
opened by the electric field.}
\label{Fig3}
\end{figure}

For the particular case of a $(1,1)$ CNT the energy spectrum can
be obtained in analytic form for any electron state, since the
system of equations \eqref{SE} consists of four equations only. 
This system results in a biquadratic equation 
for the eigenvalues $\varepsilon_E(k_s)$:
\begin{equation}\label{biquad}
\varepsilon_E^4(k_s)-\varepsilon_E^2(k_s)(w_1^2+w_2^2+2v_1+2v_2)
+(v_2-v_1+w_1w_2)^2=0,
\end{equation}
where $w_1=\gamma_0[1+2\cos(k_sa)]$,
$w_2=\gamma_0[1-2\cos(k_sa)]$, $v_1=[V\cos(\phi+\pi/3)]^2$,
$v_2=[\sqrt{3}V\sin(\phi+\pi/3)]^2$, and $V=\sqrt{3}eEa/(4\pi)$.
The energy spectrum $\varepsilon_E(k_s)$ obtained from 
Eq.~\eqref{biquad} is shown in Fig.~\ref{Fig3} (solid lines) for 
a range of wave vectors $-\pi/a\leq k_s\leq\pi/a$. 
In the Figure, positive energies correspond to the conduction 
band and negative energies to the valence band. The energy spectrum 
in the absence of the field is shown for comparison (dashed lines).
According to Eq.~\eqref{lzero}, the superlattice period $l_0$ for
a $(1,1)$ CNT is equal to twice the lattice constant $a$.
Therefore, as can be seen in Fig.~\ref{Fig3}, the width of the first
Brillouin zone in the presence of a transverse electric
field is half that without the field. It can also be seen that the
electric field opens gaps in the dispersion curve at $k_s=\pm\pi/(2a)$
due to the aforementioned Bragg reflection of electron waves.
For electric fields satisfying condition \eqref{Eless}, we obtain
from Eq.~\eqref{biquad} the Bragg gap
\begin{equation}\label{deltaeps}
\Delta\varepsilon= \frac{\sqrt{3}eEa}{2\pi}|\cos(\phi+\pi/3)|.
\end{equation}
The result in Eq.~\eqref{deltaeps} can also be obtained from the 
more general formula \eqref{Bragggeneral}. It should be noted that 
the Bragg gap, as well as the whole energy spectrum of the CNT in 
a transverse electric field, depends on the orientation of the CNT 
relative to the field (i.e. on the angle of rotation $\phi$). 
In particular, when $\phi=\pi/6$ the Bragg gap \eqref{deltaeps} is
zero: for this angle the values of the electric field potential 
at atoms $A$ and $B$ in a $(1,1)$ CNT are equal in magnitude but 
opposite in sign, and so the mean value of the potential within
one elementary cell of the CNT is zero.

In the general case of a $(n,1)$ nanotube, for external electric 
field intensities attainable in experiment ($E\sim 10^5$ V/cm) and 
for a typical nanotube of radius $R\sim 10$ {\AA}, the value of the 
Bragg gap given by \eqref{Bragggeneral} is 
$\Delta\varepsilon\sim 10^{-2}$ eV,
which is comparable to the characteristic energy of band
splitting in conventional semiconductor superlattices. As a
consequence, the discussed superlattice effects generated by the
transverse electric field in $(n,1)$ CNTs should be observable in
experiments, and may take place in existing CNT field-effect
devices.\cite{Appenzeller} The inherent regularity of a 
nanotube-based superlattice, with the superlattice period determined 
by the CNT radius, presents a distinct advantage over semiconductor
superlattices, in which monolayer fluctuations are unavoidable.
A whole range of new nanoelectronic devices based on the discussed
superlattice properties of $(n,1)$ CNTs can be envisaged, including 
Bloch oscillators\cite{Esaki} and quantum cascade lasers.\cite{Faist} 
An evaluation of the feasibility of these novel devices and selection 
of their optimal parameters will undoubtedly form the subject of 
extensive future research.

\section{Conclusions}

In this paper we have discussed a previously overlooked class of
CNTs, which may be termed `helical' nanotubes. While we have 
concentrated on the superlattice behavior of such nanotubes in a 
transverse electric field, we also expect their unique symmetry to 
manifest itself in modification of the electron-electron, 
electron-phonon and electron-photon interactions. In addition,
we have shown that superlattice behavior in a transverse electric
field is a generic feature of helical quasi-one-dimensional 
nanostructures, which raises new possibilities for developing 
optoelectronic devices operating in the terahertz range of 
frequencies. 

This work is supported by the Royal Society, INTAS, the Russian
Foundation for Basic Research, and the `Russian Universities'
program.

\end{document}